\documentstyle[12pt,epsfig]{article}

\newcommand{\beq}{\begin{equation}}
\newcommand{\eeq}{\end{equation}}
\newcommand{\bea}{\begin{eqnarray}}
\newcommand{\eea}{\end{eqnarray}}
\newcommand{\nn}{\nonumber}

\newcommand{\gtrsim}{\ \rlap{\raise 
2pt\hbox{$>$}}{\lower 2pt \hbox{$\sim$}}\ }
\newcommand{\lessim}{\ \rlap{\raise 
2pt\hbox{$<$}}{\lower 2pt \hbox{$\sim$}}\ }

\newcommand{\np}[1]{Nucl. Phys. {\bf #1}}
\newcommand{\pl}[1]{Phys. Lett. {\bf #1}}
\newcommand{\pr}[1]{Phys. Rev. {\bf #1}}
\newcommand{\prl}[1]{Phys. Rev. Lett. {\bf #1}}

\makeatletter
\if@twoside
\else \oddsidemargin 00pt \evensidemargin 00pt
\fi
\let\@eqnsel = \hfil

\expandafter\ifx\csname 
mathrm\endcsname\relax\def\mathrm#1{{\rm #1}}\fi
\@ifundefined{mathrm}{\def\mathrm#1{{\rm #1}}}{\relax}

\makeatother

\begin{document}

\title{\vskip-2.5truecm{\hfill \baselineskip 14pt {{
\small  \\
\hfill MZ-TH/99-09 \\ 
\hfill April 99}}\vskip .9truecm}
 {\bf Oscillations, Neutrino Masses and Scales of New Physics
}}

\vspace{5cm}

\author{Gabriela Barenboim\footnote{\tt 
gabriela@thep.physik.uni-mainz.de} 
\phantom{.}and Florian Scheck\footnote{\tt 
Scheck@dipmza.physik.uni-mainz.de}
 \\  \  \\
{\it  Institut f\H ur Physik - Theoretische 
Elementarteilchenphysik }\\
{\it Johannes Gutenberg-Universit\H at, D-55099 Mainz, 
Germany}
\\
}

\date{}
\maketitle
\vfill

\begin{abstract}
\baselineskip 20pt
We show that all the available experimental 
information involving neutrinos
can be accounted for within the framework of already existing models
where neutrinos have zero mass at tree level, but obtain a small
Dirac mass by radiative corrections.
\end{abstract}
\vfill
\thispagestyle{empty}

\newpage
\pagestyle{plain}
\setcounter{page}{1}

The results of the Super Kamiokande collaboration convincingly proved that
an initial beam of atmospheric muon neutrinos oscillates into other
neutrino species. As a nontrivial mass sector for neutrinos is definitely
outside of what the minimal standard model can predict, at tree level
or through radiative corrections, these results provide a
first clear evidence for the existence of new physics beyond the
minimal standard model. In this note we address the question whether
the two rather different mass differences which are needed to explain
all neutrino anomalies simultaneously, the Super Kamiokande results,
the solar neutrino deficit, and the LSND data, can be understood in a
natural manner within mild and plausible extensions of the minimal
model. We point out that this is indeed possible with implications
which, though simple, have far reaching and testable consequences. 

As is well known, in the minimal version of the standard model, the
left-handed neutrino is the only member of each family which does not
have a right-handed partner. The model makes a clear distinction between 
left and right in the classification of leptons and, therefore, in their
interactions with the gauge particles. Indeed, with left-chiral
fields in doublets, and right-chiral fields in singlets with respect
to $SU(2)_L$, charged current interactions show  maximal
violation of parity and charge conjugation. Regarding neutrino mass
terms there are then two choices: either right-chiral neutrinos are
part of the particle spectrum of the model, with all $SU(2)_L$ and
$U(1)_Y$ quantum numbers vanishing so that they do not interact with
the gauge particles, and they appear in the mass matrix due to physics
beyond the minimal model. In this case it is natural to expect neutrino
masses to be of Dirac type and to find a close relationship between
mass matrices and family mixing. 

The other possibility is that right-chiral neutrino fields do not
appear in the model at all. In this case only Majorana masses are
possible. For example, as shown in \cite{wein}, effects of quantum
gravity could generate lepton number violating, non-renormalizable 
operators of the form (using the usual short-hand notation)
\bea
\lambda_L \, \frac{ L  L H  H }{M_{\mbox{Planck}}} \;\; + \;\; 
\mbox{h.c.} \nn
\eea
which in turn would be induced by a Majorana mass term of the kind
$\mu \;  \nu_L^T \;  C^{-1}\; \nu_L$.
Here $M_{\mbox{Planck}} \approx 1.22 \cdot 10^{19}$~GeV denotes the Planck
mass, while $\lambda_L$ is an effective dimensionless coupling
constant that one would expect to be of order one.
Such a mass would lead to squared mass differences of the order of
$10^{-9}$~eV$^2$, by far too small to explain any of the existing
asymmetries, even with ``anomalous'' fine tuning of $\lambda_L$. Of
course, one might ask whether the mass scale in the denominator could be
chosen differently, for example, by taking it to be the grand
unification scale $ M_{\mbox{\footnotesize GUT}} \approx 2 \cdot 10^{16}$~GeV.
Unfortunately, this not only would yield masses which are still too low, but 
also would call by itself for new physics beyond the standard model.
(What other reason would there be to introduce a new scale?) Thus, it
appears that the obesrved neutrino anomalies cannot be accommodated within 
the minimal standard model in a natural manner, even with inclusion of
effects from quantum gravity, and, therefore, the need arises to
extend the model such that neutrino physics can be understood in a
framework which already contains the other fermions, scalars and gauge
bosons (the known ones and, hopefully, the ones to be discovered soon).

In this situation and with neutrino oscillations being positively
established the challenge for theorists is to explain why the masses
of neutrinos are small, as compared to the ones of their charged
partners, and to find out whether or not there is violation of total
lepton number
\bea
L := L_e + L_\mu + L_\tau\; . 
\eea
As is well known Dirac masses are compatible with $L$-conservation
while Majorana masses imply $\Delta L=\pm 2$.

Many extensions of the standard model assume neutrinos to be Majorana
particles and understand the smallness of their mass in terms of the 
see-saw mechanism. The underlying idea is quite simple: In a basis of
Majorana fields\footnote{We remind the reader that in studying spinor
  representations of the Lorentz group, for fermions which do not
  carry any additively conserved quantum number, one naturally finds
  Majorana spinors $\Psi_M^{(i)}$ rather than Dirac spinors. Only when 
  two Majorana fields happen to have the same mass can one form Dirac
  fields such as $\Psi_D=\Psi_M^{(1)}+i\Psi_M^{(2)}$. From this point
  of view one might say that Majorana fields are more fundamental
  than Dirac fields.} the 
mass matrix has the general form
\begin{equation}\label{MDmass}
  M = \left(
    \begin{array}{cc}
\mu_M^{(l)} & m_D \\ m_D & \mu_M^{(h)}
    \end{array}\right)\, .
\end{equation}
Here, when dealing with one family, $\mu_M^{(l)}$ and $\mu_M^{(h)}$ are
Majorana mass terms which may be chosen real and, say, positive, $m_D$ 
is a Dirac term and may still be relatively complex,
$m_D=\mu_De^{i\phi}$. In the case of three families these quantities
are corresponding block matrices. It seems natural to assume $\mu_D$
to be of the order of the charged partner's mass, $\mu_D\approx
m(l)$, and, from the smallness of physical neutrino masses, to
assume the lower Majorana term to fulfill $\mu_M^{(l)}\ll\mu_D$. In contrast
to the former, the second Majorana term is expected to be large as
compared to the charged partner's mass, $\mu_M^{(h)}\gg\mu_D$, and to 
represent the signal of a new mass scale. The eigenstates of
$M$, eq.~(\ref{MDmass}), are Majorana fields, the corresponding
eigenvalues being approximately \cite{scheckb} 
\begin{displaymath}
  m_1\approx \frac{\mu_D^2}{\mu_M^{(h)}}\; ,\qquad
  m_2\approx \mu_M^{(h)}\, .
\end{displaymath}
The first of these is small and proportional to the \emph{square} of the
charged partner's mass, while the second is large and proportional to
the new mass scale\footnote{We note in passing that if the two
  Majorana terms were equal and if $\phi$ were an odd multiple of $\pi
  /2$ the eigenvalues of (\ref{MDmass}) would be equal. In this case a
  Dirac state is formed.}. 
Since the heavier neutrino state is absent in the standard model and
has not been observed experimentally, it is plausible to assume that
the new scale is associated with either a local \cite{loc} or global
\cite{glo} $B-L$ symmetry. As this mass must be in the range of a TeV
or higher, a natural explanation for the smallness of neutrino masses is
achieved. However, the proportionality to the squared charged lepton
masses, $m(\nu_i)\propto m^2(l_i)$, leads to difficulties in trying to 
fit all neutrino anomalies simultaneously. Typically, these models
contain two very different mass differences. Therefore, if one wishes
to explain the atmospheric and LSND neutrino data with the same mass
difference (as should be done if the three anomalies are to be
understood within three flavours, i.e. with two mass differences only),
the remaining mass difference is too small to account for the vacuum
oscillation solution of the solar neutrino data while being too 
large for the MSW effect to be operative.

Let us then turn to the possibility that neutrinos are Dirac
particles, not Majorana, and that the leptonic mass matrices
conserve total lepton number\footnote{Possible variations of the
  see-saw mechanism were proposed so as to realize this possibility
  also in a natural manner \cite{alt} by postulating the existence of
  a new scale. This time, however, the scale is not connected with any
  obvious symmetry and therefore, such models are less appealing than
  the original see-saw scheme.}. 
Indeed, an attractive possibility seems to us
that neutrinos initially are massless in a tree level Lagrangian, and
that their small (Dirac) masses are due entirely to quantum corrections.
Such ideas were discussed first in \cite{fi} and were fully developed
later, when they were studied in the context of extensions of the
minimal model, most notably left-right symmetric models \cite{lr} 
where this possibility  finds a rather economical and elegant
realization. It is the purpose of this work to embed the empirical
solutions of the neutrino anomalies found in \cite{nos} in these
models and to analyze the predictions this gives rise to. We outline
somewhat schematically the rich set of phenomenological implications
which follow from this and which are interesting in their own right.

Going backwards in time, one realizes that, though neutrinos were
considered to be Dirac particles in the original version of the 
left-right symmetric models \cite{olr}, the smallness of their mass
was not understood. Later on, a version of these models was developed
\cite{rver} wherein neutrino masses vanish at tree level but arise at
one-loop or two-loop level, thus rendering them naturally much smaller
than the corresponding charged fermion masses. An interesting feature
of these models that makes them particularly predictive, is that
neutrino masses scale \emph{linearly} with the charged fermion
masses. As they are inherently different from the class of left right
symmetric models that lead to Majorana masses we give here a brief
sketch but refer the interested reader to the original literature for
a more detailed discussion. 

The electroweak sector of the models we consider is based on
left-right symmetry with gauge group $SU(2)_L\times SU(2)_R\times
U(1)$. The assignment of the known quarks and leptons of every
generation is as follows, 
\bea
Q_L\equiv \pmatrix{u \cr d}_L :\quad \left( 2, 1, \frac{1}{3} \right) \;\;\; ,
\;\;\;
Q_R\equiv \pmatrix{u \cr d}_R :\quad \left( 1, 2, \frac{1}{3} \right)\nn \\
\\
\Psi_L\equiv \pmatrix{\nu \cr e}_L :\quad \left( 2, 1, -1 \right) \;\;\; ,
\;\;\;
\Psi_R\equiv \pmatrix{\nu \cr e}_R :\quad \left( 1, 2, -1 \right)\nn 
\eea

At this point we depart from the minimal left-right symmetric
model, the difference being that our models also assume heavy singlet
quarks and leptons with vector-like couplings and make use of the
see-saw mechanism for quarks and charged leptons instead of
neutrinos. The idea of partial see-saw mechanism for down type quarks
and charged leptons was first discussed in \cite{lr}. Later is was
proposed \cite{pss} that the entire  fermion spectrum be generated via
the see-saw mechanism, by postulating the following heavy fermions in
the left-right symmetric model, in addition to the known fermions: two
singlet quarks denoted P and N, and a singlet lepton, denoted E, 
\bea
P:\; \left( 1, 1, \frac{4}{3}\right) \;\;, \;\; 
N:\; \left( 1, 1, - \frac{2}{3}\right) \;\;, \;\; 
E:\; \left( 1, 1, -2 \right) \;\;, \;\; 
\eea
An advantage of these models is their relatively simple Higgs structure.

For the sake of simplicity we focus on three models which
differ by their Higgs content. In the first of them, say (A), only one
pair of Higgs doublets and a parity odd real singlet scalar (needed to
generate the left-right symmetry) are needed, 
\bea
\chi_L:\; \left( 2, 1, 1\right) \;\;, \;\; 
\chi_R:\; \left( 1, 2, 1\right) \;\;, \;\; 
\sigma:\; \left( 1, 1, 0 \right) \;\;, \;\; 
\eea
In this model, neutrino masses vanish at tree level and arise only at
the two-loop level. The relevant two-loop graph is shown in 
Fig 1(a), it leads to small and finite Dirac masses given by
\bea\label{modA}
m(\nu_i) = \left( \frac{\alpha}{4 \pi \sin^2\theta_w} \right)^2
\left( \frac{ m_t m_b}{M_{W_L}^2} \right) \left(\frac{m_E^2}{M_{W_R}^2}
\right)
m(l_i)\, .
\eea
Here $m(l_i)$ is the mass of the charged lepton of the $i$-th 
generation, $m_E$ is the mass of the singlet lepton. 

In the second scenario, say (B), the following set of Higgs bosons
is chosen in order to break the gauge symmetry down to $U(1)_{em}$ and 
to give the fermions their masses
\bea
&\phi_2:\; \left( 2, 1, 0\right) & \nn \\ 
&&\nn \\
&\chi_L:\;  \left( 2, 1, 1\right) \; , \qquad
\chi_R:\;  \left( 1, 2, 1\right) &\\ 
&& \nn \\
&n_a:\;  \left( 1, 1, 0 \right)\, , \quad   a=1,2\, . &\nn
\eea
The one-loop graph which contributes to Dirac neutrino masses in this
case, is shown in Fig 1(b) and can be estimated to yield 
\bea \label{modB}
m(\nu_i) = \left( g\frac{m(l_i)}{2 M_{W_L}} \right)
\left( g\frac{ m_t }{2 M_{W_L}} \right) \frac{m_b}{16 \pi^2}
 \left(\frac{v_R}{\sigma }
\right)^3
\eea
where $\sigma$  is the vacuum expectation value (vev) of the 
singlets $n_a$, and where $v_R$ is the usual vev of the neutral member
of the right-handed doublet. Notice that the ratio of scales in this
case is somehow the inverse of the previous case:  In eq.~(\ref{modA})
the right-handed scale appears in the denominator while the
vector-like scale $m_E$ is in the numerator.

As a third model, say (C), let us introduce the following Higgs
multiplets, in addition to those of model (B),
\bea
&\phi_l:\; \left( 2, 2, 0\right) & \nn \\
&& \\ 
&n_3:\; \left( 1, 1, 0 \right)  \nn
\eea
In this model finite neutrino masses arise from the graph of Fig
1(c). Since in this case $W_L - W_R$ mixing is finite (in the previous
cases the $W_L - W_R$ mixing  vanishes at tree level and is induced
at one-loop level only), we find
\bea \label{modC}
m(\nu_i ) = \left( \frac{\alpha}{4 \pi \sin^2\theta_w} \right)^2
\left( \frac{ m_t m_b}{M_{W_R}^2}\right) m(l_i)\; .
\eea

We now turn to the phenomenological neutrino spectrum that is
required to fit all anomalies simultaneously, i.e, the deficits
observed in the solar neutrino flux and in the atmospheric neutrino
data, as well as the appearance signal of electron anti-neutrinos in
the initial $\overline{\nu_\mu}$ beam of the LSND
experiment. Obviously, in a mixing scheme involving three flavours
one has only two mass differences. In the models just described these
are given by the mass difference of the corresponding charged leptons
divided by the right-hand scale (model (C)), or multiplied by a ratio
of scales (models (A) and (B)). Therefore, once one of these mass
differences is fixed, the other is predicted. Furthermore, in fixing
any one of the two mass differences we obtain a prediction for the
scale of new physics, the two quantities being directly related by 
eq.~(\ref{modA}), eq.~(\ref{modB}), or eq.~(\ref{modC}). The new scale
can be probed through the predictions it implies, either in the lepton
sector (magnetic moment of the neutrinos, radiative decays, etc.) or
in the quark sector (e.g. sizeable CP violating rate asymmetries), or
else its existence can be disproved, for example, by an erroneous  
$K_L - K_S$ mass difference.

Following our previous work \cite{nos} we assume that at low energies
there are just three neutrino flavours and, hence, only two
differences of squared masses. In order to account for both, the
atmospheric neutrino anomaly at low energy as well as the observation
of LSND we fix the larger of these to be 
\bea \label{Delta}
\Delta M^2 = m_3^2 -m_2^2 =.3 {\hbox{ eV}}^2\; .
\eea
In the case of model (C), eq.~(\ref{modC}), this means that
\bea
m^2(\nu_\tau )  - m^2(\nu_\mu )
 = \left( \frac{\alpha}{4 \pi \sin^2\theta_w} \right)^4
\left( \frac{ m_t m_b }{M_{W_R}^2}
\right)^2 \left( m^2(\tau ) - m^2(\mu ) \right) \approx .3 {\hbox{ eV}}^2\, .
\eea
Inserting the known quantities in this expression we deduce the value
\bea \label{MWR}
M_{W_R} \approx 4{\mbox{ TeV}} 
\eea
for the right-hand scale. When we calculate the second mass difference 
from this estimate we find 
\bea
\Delta m^2 = m_2^2 -m_1^2 \approx m^2(\nu_\mu )  - m^2(\nu_e )\approx
10^{-3} {\hbox{eV}}^2 \, ,
\eea
i.e. exactly the value that we needed to explain the solar neutrino
deficit \cite{nos}. Notice that $\Delta m^2$ is an output of our model
and that no additional input was used to fix its value. Such a mass
difference is also called to play an important role in the atmospheric
neutrino anomaly for upward going events \cite{nos2}.

The value (\ref{MWR}) for the right hand scale which is implied by our 
model is high enough to evade the stringent bounds from the $K_L -
K_S$ mass difference. On the other hand, it leads to strong
predictions which should be observable in ongoing and future
experiments. For example, at the forthcoming B-factories  
the CP violating rate asymmetries should show clear deviations from
the values expected within the minimal standard model.

Regarding the other two scenarios, (A) and (B), only the ratio of the
different scales involved can be predicted and we cannot obtain as definite 
predictions as those we have obtained before, in model (C). However,
we still can vary these ratios within reasonable limits and can
explore the range of admissible masses.

Let us begin with model (A). The experimental lower bound on the mass
of the heavy charged lepton from direct searches, $M_E > 42.8 $ GeV,
automatically places a lower bound on the right-handed gauge boson mass
\bea
M_{W_R} > 2.2 {\mbox{ TeV}}\; .
\eea
This bound ensures that the new contributions to flavour changing neutral 
current processes such as $K_L \longrightarrow \mu {\overline{\mu}}$
and the $K_L -K_S $ mass difference are well within present experimental
limits. Even more, the range of masses of the heavy charged lepton
required in order to have a non-decoupled right-handed gauge boson is
within what is being tested now and what will be tested in the near
future in accelerator experiments. As can be seen from Fig 2, only
masses in the range $m_E=43$ - $420$ GeV can give phenomenologically
interesting right-handed gauge bosons, of masses lighter than 20
TeV. Therefore if model (A) has some truth in it this will soon be
seen either by an observation of this relatively light charged lepton
$E$ (and the associated heavy vector-like quarks), or by indirect
effects of the right-handed gauge bosons. 

In model (B), the quantity $\sigma$ is again associated with the 
mass of the heavy charged lepton, which is $ M = f \sigma$ with 
$f$ a coupling constant presumably of order one. As the smallness of
the masses of the down type quarks is explained in this model by a
see-saw type mechanism 
\bea
m_{q_d} \propto \frac{v_L v_R}{\sigma}\, ,
\eea
$v_R$ is expected to be much smaller than $\sigma$. This leads to a
pattern of masses which is just the inverse of the previous one,
i.e. now the heavy lepton masses are much larger than the mass of the
right-handed gauge boson.
For the same range of phenomenologically acceptable masses of the
right-handed gauge bosons we have used before, 1.6 TeV $\leq M_{W_R}
\leq $ 20 TeV, the mass of the heavy charged lepton is now in an
interval spanning from 195 TeV to 2440 TeV, cf. Fig. 3. Thus, we have
a model where the vector-like quarks are almost completely decoupled. 
Therefore, unlike the previous case, this model resembles the original
left-right model in the sense that the low energy phenomenology
associated with the enlargement of the fermion and Higgs sectors is 
reduced to radiatively induced Dirac masses of neutrinos only.

In summary we show that it is possible to explain all neutrino
observations within the framework of previously developed models 
where neutrinos have zero masses at tree level, but obtain small Dirac
masses by radiative corrections. Phenomenological patterns of neutrino
masses consistent with the observed anomalies were proposed
previously, their sole purpose being to explain the anomalies, but, to
the best of our knowledge, they were not part of any
complete model. We do not know of any model accounting for all available
experimental information involving neutrinos where the mass differences
are somehow justified by an underlying principle. As far as we know, our
models provide the first attempt to do so. The choice to describe the
light neutrinos as Dirac particles, as opposed to Majorana particles,
is largely dictated by lepton number conservation, which seems
to be conserved in nature. By the same token, this choice puts
all fermions on equal footing as far as their additive quantum numbers 
and their masses are concerned.

To be specific we made use of three simple extensions, known since long,
of the left-right symmetric model in which neutrinos are Dirac particles,
to account for all experimental information revealed by the observed
neutrino oscillations. The masses of our neutrinos arise either at the
one-loop or the two-loop level, thus explaining why they are small as 
compared to the masses of their charged partners. In some
variations, the scale for new physics that has to be introduced,
(i.e. the $W_R$ and the vector-like fermion mass) could be light
enough to be probed in the next generation of experiments.

We emphasize again the importance of avoiding any ad-hoc
assumptions. Roughly speaking, instead of introducing model
assumptions designed to deal with the neutrino anomalies we proceed in 
the opposite way. By fixing one mass difference, within a specific
model and not meant to explain any neutrino deficit or appearance, the
remaining difference as well as the new scales are obtained as
definite predictions. This observation illustrates that, despite the
danger of a phenomenological disaster (such as an unacceptably low
$W_R$ mass, or flavour-changing neutral currents that are too
large), there are models in which the problems raised by neutrinos can
be solved in a consistent way and without ad-hoc assumptions. At the
same time such models have very interesting phenomenological
implications to which we hope to return in the near future.

\vspace{.5cm}

\begin{center}
{\bf Acknowledgements}
\end{center}

We are very grateful to Oscar Vives
for  enlightening
discussions. A post-doctoral fellowship
of the Graduiertenkolleg ``Elementarteilchenphysik bei 
mittleren und hohen Energien"
of the University of Mainz is also acknowledged.

\newpage

\begin{figure}[!ht]
  \begin{center}
  \epsfig{file=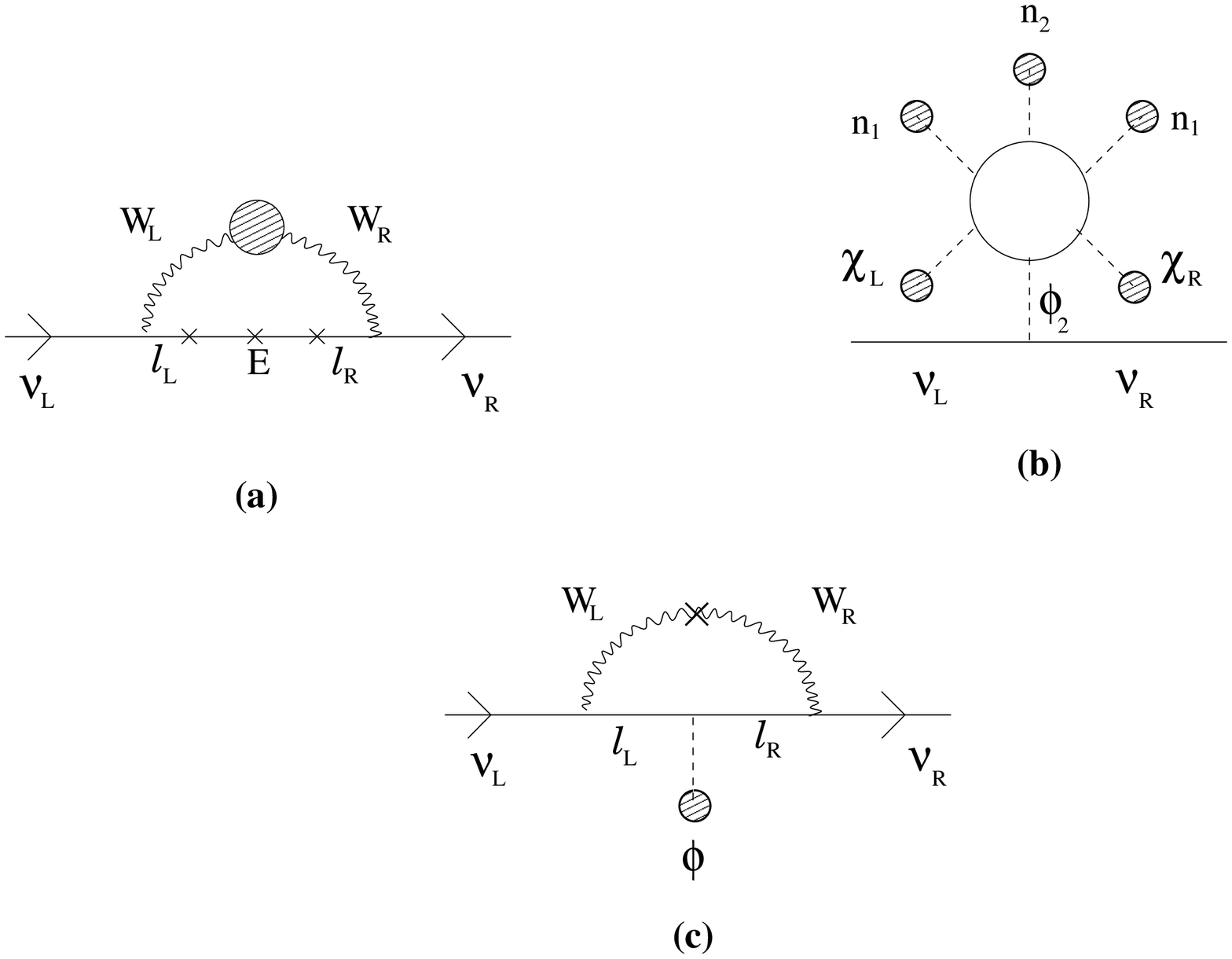,width=15cm}
\caption{Radiatively generated Dirac masses for the different models} 
  \end{center}
\end{figure}

\pagebreak

\begin{figure}[!ht]
  \begin{center}
  \epsfig{file=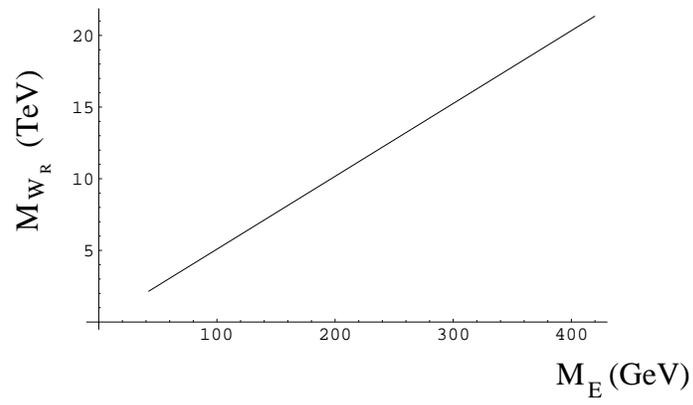,width=9cm}
  \parbox{15cm}{\caption{Right handed gauge boson mass as a function
of the heavy charged lepton mass}}
  \end{center}
\end{figure}

\begin{figure}[!ht]
  \begin{center}
  \epsfig{file=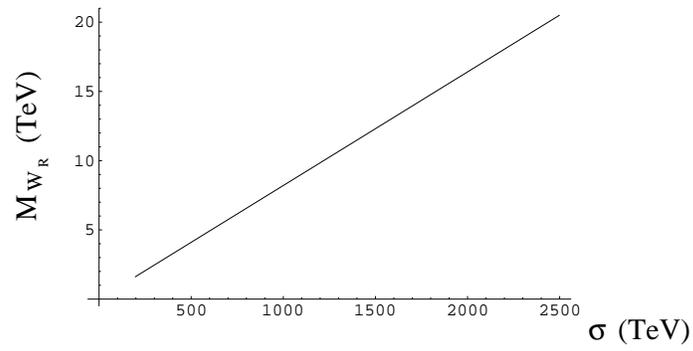,width=9cm}
  \parbox{15cm}{\caption{Right handed gauge boson mass as a function
of the singlet's vev}}
  \end{center}
\end{figure}

\end{document}